\documentclass[aps,prl,groupedaddress,twocolumn,floatfix]{revtex4-1}

\usepackage{mathrsfs}
\usepackage{xcolor}
\usepackage{epsfig}
\usepackage{relsize}
\usepackage{graphicx}
\usepackage{amsfonts}
\usepackage[figuresright]{rotating}
\usepackage{amssymb}
\usepackage{amsmath}
\usepackage{color}
\usepackage{graphicx,bm,units,yfonts}
\usepackage{mathrsfs}

\DeclareGraphicsExtensions{.eps, .PNG, .jpg}

\newcommand{\ket}[1]{| #1 \rangle}
\newcommand{\bra}[1]{\langle #1 |}

\begin{document}

\title{Quantized frequency-domain polarization of driven phases of matter}
\author{Ian Mondragon-Shem,$^1$ Ivar Martin,$^2$  A. Alexandradinata,$^1$  and  Meng Cheng$^1$}
\affiliation{$^1$ Department of Physics, Yale University, New Haven, Connecticut 06520, USA\\
$^2$Materials Science Division, Argonne National Laboratory, Argonne, Illinois 60439, USA}
\date{\today}
\begin{abstract}
Periodically driven quantum systems can realize novel phases of matter that do not exist in static settings. We study signatures of these drive-induced phases on the $(d+1)$-dimensional Floquet lattice, comprised of $d$ spatial dimensions plus the frequency domain. The average position of Floquet eigenstates along the frequency axis can be written in terms of a non-adiabatic Berry phase, which we interpret as frequency-domain polarization. We argue that whenever this polarization is quantized to a nontrivial value, the phase of matter cannot be continuously connected to a time-independent state and, as a consequence, it captures robust properties of its dynamics. We illustrate this in driven topological phases, such as superconducting wires and the anomalous Floquet Anderson insulator; as well as in driven symmetry-broken phases, such as time crystals. We further introduce a new dynamical phase of matter that we construct by imposing quantization conditions on its frequency-domain polarization. This illustrates the potential for using this kind of polarization as a tool to search for new driven phases of matter. 
\end{abstract}

\maketitle

\textit{Introduction.} A long-standing interest in condensed matter physics has been to understand the properties of non-equilibrium quantum systems. Over the past decade, significant attention has been given to phases of matter induced by periodic drives, known as Floquet phases. This has led, for example, to the discovery of new topological states \cite{Jiang2011,Titum2015,Karzig2015,Titum2017,Rudner2013, Potter2016, Roy2017, Harper2017,SYao2017}, which exhibit novel phenomena such as non-adiabatic charge pumping and quantized magnetization \cite{Titum2016,Lindner2017}. New symmetry-broken phases have also been discovered, such as driven spin-glasses that spontaneously break time-translation symmetry  \cite{vK2016b, VKhemani2016,Yao2017,Choi2017}. 

The standard approach to Floquet phases has been to analyze their time-dependent properties within a single period, the so-called micromotion. There has been, however, growing interest in understanding Floquet systems in the frequency domain. Traditionally, the frequency domain has been used as an auxiliary tool to simplify numerical calculations and make useful analytical approximations \cite{Rudner2013, Eckardt2015,Rodriguez2018}.  Recent studies have led to interesting insight into some aspects of the dynamics of Floquet states on the Floquet lattice, which is composed of spatial and frequency directions \cite{Gomez2013, Ivar2017, Baum2018, Peng2018a, Crowley2018}. There is still lacking, however, an understanding of the signatures of drive-induced phases that arise in the frequency domain. Since a frequency domain description naturally captures the full micromotion of a system, it is tantalizing to look for properties on the Floquet lattice that capture robust aspects of their dynamics.

In the present work, we show that drive-induced phases of matter are characterized by non-adiabatic Berry phases which can be interpreted as polarization along the frequency axis of the Floquet lattice. We argue that when a phase of matter exhibits a nontrivial quantized frequency-domain polarization, it cannot be continuously connected to a static state. This kind of polarization thus captures robust features of the dynamics of drive-induced phases. We present examples in 1D and 2D of quantized frequency-domain polarization as well as quantized frequency-domain pumping. Finally, using these insights, we discuss how frequency-domain polarization can be used to construct new dynamical phases of matter.

\textit{Frequency-domain polarization as a non-adiabatic Berry phase.}  We are interested in the physics of a $d$-dimensional system described by a time-periodic Hamiltonian $h(t)=h(t+T).$ The conventional approach to study this problem is to compute the set of Floquet eigenstates $\{\ket{u^{(n)}_{t}}\}$ which satisfy
\begin{eqnarray}
\mathcal{U}(t+T,t)\ket{u^{(n)}_{t}}=e^{-i\epsilon_n T}\ket{u^{(n)}_{t}}, \label{Eq_Flt}
\end{eqnarray}
where $\mathcal{U}(t,t_0)=\mathcal{T}e^{-i \int^t_{t_0} d\tau h(\tau)},$ with $\mathcal{T}$ the time ordering operator.  The set $\{\epsilon_n\}$ is referred to as the quasi-energy spectrum, taking values in the range $(-\frac{\pi}{T},\frac{\pi}{T}].$  By studying the time-dependent properties of $\{\ket{u^{(n)}_{t}}\},$ new phases of matter have been found that do not exist in time-independent systems.

In contrast to this approach, we seek to understand driven phases in the frequency domain. We will set the unit frequency to  $\omega_0=2\pi/T=1.$ We begin by writing the Schrodinger equation as $\left[-i\partial_t+h(t)\right]\ket{v^{(n)}_{t}}=\epsilon_n \ket{v^{(n)}_t},$  where $\ket{v^{(n)}_t}=e^{i\epsilon_n t}\ket{u^{(n)}_t}$ is periodic in $t.$  This eigenvalue equation can be viewed as describing a $(d+1)$-dimensional system with a Hilbert space $\mathscr{F}=\mathscr{H}\otimes \mathscr{C},$ where $\mathscr{H}$ is the space of the static system and $\mathscr{C}$ is the space of time-dependent functions of period $2\pi$ \cite{Rodriguez2018,Eckardt2015}. In $\mathscr{C},$ temporal  and frequency bases $\{\ket{t}\}$ and $\{\ket{m}\}$ can be defined such that $\langle n\ket{m} =\delta_{nm},$ $\langle t\ket{t'} = 2\pi \delta(t-t') ,$ and $\langle t\ket{m}=e^{i m t}.$  The Schrodinger equation in $\mathscr{H}$ is then the temporal representation of the eigenvalue problem in $\mathscr{F}$ given by \cite{Rodriguez2018}
\begin{eqnarray}
\left[\hat{\omega}+\mathcal{H}\right]\ket{\Psi^{(n)}_M}=\left(M+\epsilon_n\right)\ket{\Psi^{(n)}_M},\label{Eq_FF}
\end{eqnarray}
where $\hat{\omega}=\sum_m m\, \mathbb{I}\otimes\ket{m}\bra{m} $ is the discrete frequency operator with $\mathbb{I}$ the identity in $\mathscr{H}$, and  $\mathcal{H}=\frac{1}{2\pi}\sum_{mm'}\int_0^{2\pi}d\tau e^{i(m-m')\tau}h(\tau)\otimes \ket{m}\bra{m'}$ is the Hamiltonian in $\mathscr{F}.$ The integer $M$ labels solutions $\ket{\Psi^{(n)}_M}=\frac{1}{2\pi}\int^{2\pi}_0dt e^{i M t}\ket{v^{(n)}_t}\ket{t}$ that correspond to the same time-dependent Floquet state.

To understand the physics of the driven problem we must thus understand the states $\{\ket{\Psi^{(n)}_{M}}\}$ on the Floquet lattice. It is insightful in this regard to view Eq.\ref{Eq_FF} as an analogue of a Wannier-Stark ladder \cite{Wannier1960,Wannier1962}, in which the frequency operator takes the place of the position operator. In particular, the average position of an electron in the Wannier-Stark ladder is given by an adiabatic Berry phase \cite{Zak1989}. Interestingly, in the Floquet problem the average frequency is also given by a Berry phase \cite{Moore1991}
\begin{eqnarray}
\bra{\Psi^{(n)}_{M}}\hat{\omega}\ket{\Psi^{(n)}_{M}}&=&M+\mathcal{P}^{\omega}_n,\label{Eq_wb}
\end{eqnarray}
where $\mathcal{P}^{\omega}_n= \frac{1}{2\pi}\int^{2\pi}_0 dt\mathcal{A}^{nn}_t(t),$ and $\mathcal{A}^{nm}_t(t)=-\bra{v^{(n)}_t} i\partial_t \ket{v^{(m)}_t}.$  The Berry phase $ 2\pi \mathcal{P}^{\omega}_n,$ however, is non-adiabatic in nature because it is not calculated using the instantaneous eigenstates of the Hamiltonian but, instead, it is calculated using Floquet eigenstates. It is revealing that the quasi-energies can be written as $\epsilon_n=\overline{E}_n+\mathcal{P}^{\omega}_n,$ where $\overline{E}_n=\frac{1}{2\pi}\int_0^{2\pi}d\tau\bra{u^{(n)}_{\tau}}\mathcal{H}(\tau)\ket{u^{(n)}_{\tau}}$ is the average energy per cycle. Thus, the quasi-energies are not entirely dynamical quantities, but also contain geometric information.  

Although the Berry phase $ 2\pi \mathcal{P}^{\omega}_n$ itself has been studied previously \cite{Aharonov1987,Moore1990a, Moore1990b}, the physics underlying its relation with the average of $\hat{\omega}$ has not received as much attention, as far as we are aware. In static systems, the modern theory of polarization hinges on an analogous relation $\bra{\Psi^{(n)}_{R}}\hat{x}\ket{\Psi^{(n)}_{R}}=R+\mathcal{P}^{x}_n,$
where $\hat{x}$ is the position operator, $\ket{\Psi^{(n)}_{R}}$ are exponentially localized Wannier functions in position space, $R$ is a Bravais-lattice vector, and $\mathcal{P}^{x}_n= \frac{1}{2\pi}\int^{2\pi}_0 dk\mathcal{A}^{nn}_k(k),$ with $\mathcal{A}^{nn}_k(k)$ the Berry connection in quasimomentum space \cite{Marzari1997}. Since $\mathcal{P}^{x}_n$ measures deviations with respect to integer positions, it quantifies spatial polarization.  Thus, by analogy, we are motivated to interpret $\mathcal{P}^{\omega}_n$ as polarization along the frequency direction. 

Let us examine how this kindf of polarization can carry signatures of drive-induced phases. To begin with, all static phases must have vanishing frequency-domain polarization. This follows from $\mathcal{H}$ being diagonal in frequency space in the absence of a drive, which implies that $\ket{\Psi^{(n)}_M}$ is completely localized to single sites along the frequency axis. When a drive is turned on continuously, static phases can become polarized in frequency space due to the induced time dependence. However, as long as no phase transition is crossed, this polarization can always be removed continuously by turning off the drive. Thus, a nonzero $\mathcal{P}^{\omega}_n$ is not in itself a signature of a drive-induced phase. By contrast, if a driven system has a nontrivial {\em quantized} frequency-domain polarization, it will not be possible to remove the drive continuously. As a result, such a phase is not continuously connected to a static system, making it inherently dynamical in nature. Frequency-domain polarization can thus serve to characterize robust aspects of the dynamics of drive-induced phases, as we will illustrate in the examples that follow. 

Keep in mind that for  $\bra{\Psi^{(n)}_{M}}\hat{\omega}\ket{\Psi^{(n)}_{M}}=\sum_m m\vert \Psi^{(n)}_M(m)\vert^2$ to be well-defined, the probability density $\vert \Psi^{(n)}_M(m)\vert^2$ must be sufficiently localized along the frequency direction. This will be the case, for example, if the drive is an analytic function of time \cite{Marzari1997, Emin1987}. We have found, however, that numerical convergence is also achieved in the models with discontinuous time dependence which we investigate here.

\begin{figure}
\includegraphics[trim = 0cm 0.cm 0cm 0cm, clip,scale=0.18]{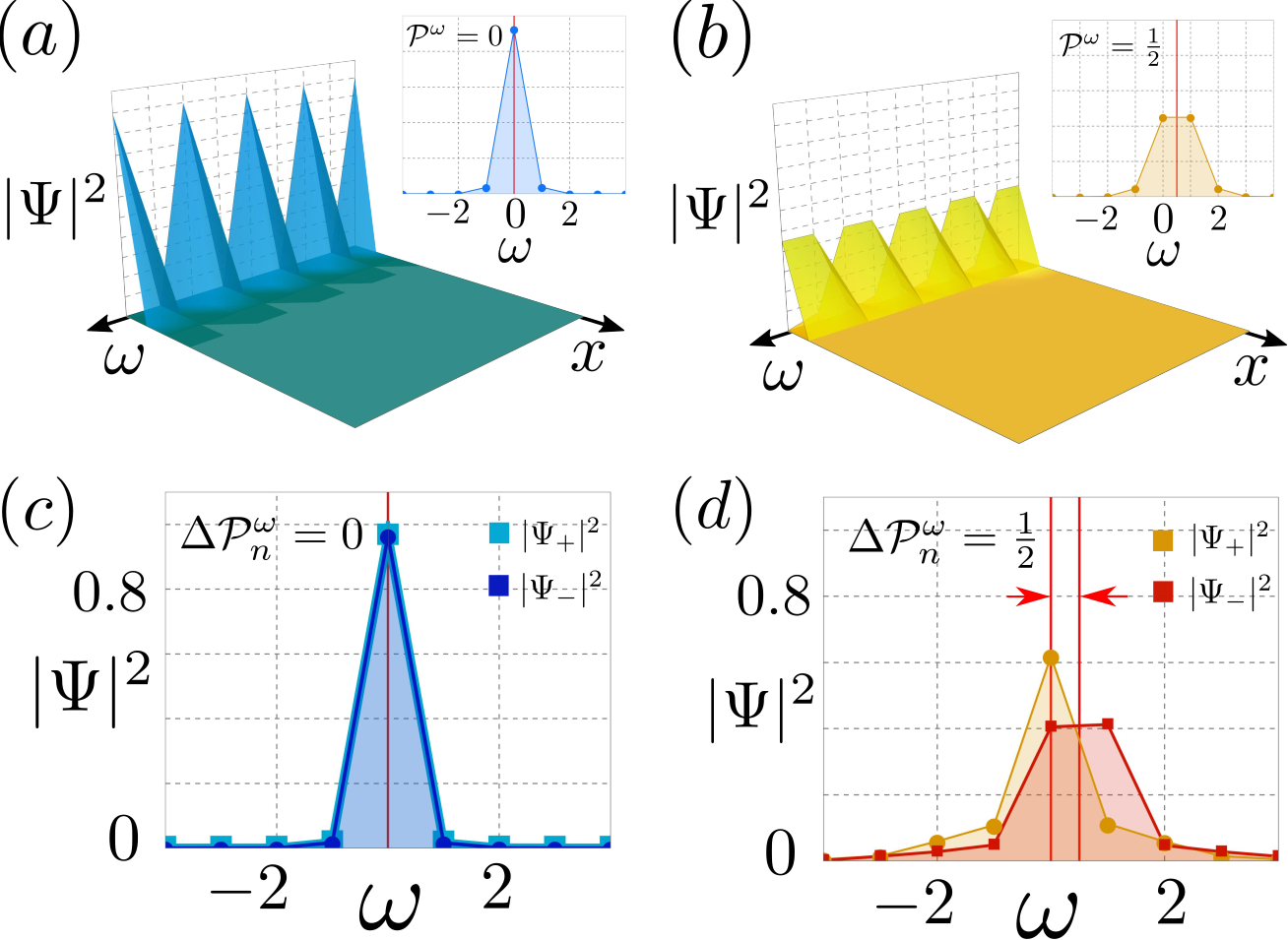}
\caption{\textbf{(a,b)} Probability densities of $0$ and $\pi$-modes along one edge of the Floquet lattice with $\{\Gamma,\overline{J_r}\}=\{0, 1/4\}$ and $\{\Gamma,\overline{J_r}\}=\{1/2, 1/4\},$ respectively. The insets show the density profiles, and the red lines indicate their polarization. \textbf{(c,d)} Probability densities of the spectrally paired states of the two kinds of spin glasses of the spin model with the same parameters as in (a,b). The quantized relative frequency-domain polarization in the time-crystal phase \textbf{(c)} is noted with the red arrows.}\label{Fig_Majorana}
\end{figure}

\textit{Quantized frequency-domain polarization.} An example of quantized frequency polarization arises at the boundary of driven one-dimensional spinless superconductors (SC), which reside on a 2D Floquet lattice. We will consider the non-interacting case for simplicity which can be described by a mean-field Hamiltonian in the Bogoliubov-de Gennes (BdG) formalism.  Static SCs can realize a topological phase with $0$-energy Majorana boundary modes. When the system is driven, a new Majorana mode can appear at $\epsilon=1/2,$ which is referred to as $\pi$-mode. Due to the particle-hole symmetry of the BdG Hamiltonian, the average energy per cycle of both types of boundary modes vanishes $\overline{E}_\eta=0$ ($\eta=0,\pi$). Thus, we obtain
\begin{eqnarray}
\mathcal{P}^{\omega}_{\text{edge},\eta}&=&\frac{1}{2\pi}\eta. \label{Eq_TBPE}
\end{eqnarray}
This implies that driven 1D SCs are characterized by quantized polarization at the edge of the 2D Floquet lattice. We illustrate this with the model $h(0<t<\pi)=\sum_r  \Gamma i a_r b_{r}$ and $h(\pi<t<2\pi)=\sum_r  J_r ib_r a_{r+1}$,
where $a_r=c^{\dagger}_r+c_r$ and $b_r=i(c^{\dagger}_r-c_r)$ are Majorana fermion operators. In  Fig.\ref{Fig_Majorana}a,b we show the probability densities of both types of Majorana modes on the Floquet lattice. Changing the parameters of the Hamiltonian continuously can change these densities, leading to non-universal changes in the their dynamics. However, the nonzero polarization of the $\pi$ modes makes it impossible to take away their time-dependence, thus revealing the inherent dynamical nature of the associated Floquet phase.

This example can be used to show that frequency domain polarization also characterizes symmetry broken phases. If we perform a Jordan-Wigner transformation \cite{Fradkin2013}, the SC wire maps to the spin-$1/2$ model $h(0<t<\pi)=\sum_r \Gamma \sigma^1_r$ and $h(\pi<t<2\pi)=\sum_r J_r\sigma^3_r \sigma^3_{r+1}.$ The SC phase with $0$-modes maps to a phase that exhibits spin-glass order which breaks the discrete $\mathbb{Z}_2$ symmetry $P=\prod_r \sigma^1_r$ \cite{vK2016b}. In this phase, the Floquet states that differ by the occupation of the $0$-mode in the fermion language correspond to doubly degenerate spin states $\ket{u^{(n)}_{t,\pm}}$ with quasi-energies $\epsilon^{(n)}_+=\epsilon^{(n)}_-,$ where $\pm$ labels eigenvalues of $P.$ Since the average energy per cycle must be the same for both states, the difference in their frequency-domain polarization vanishes $\Delta \mathcal{P}^{\omega}\equiv\mathcal{P}^{\omega}_{n,+}-\mathcal{P}^{\omega}_{n,-}=0.$  In contrast, the SC phase with $\pi$-mode maps to a phase which  exhibits spin-glass order and spontaneously breaks time-translation symmetry. In this case, the spin states $\ket{u^{(n)}_{t,\pm}}$ that differ by the occupation of the $\pi$-mode in the fermion language are spectrally separated $\epsilon^{(n)}_+-\epsilon^{(n)}_-=1/2,$  which has been shown as the root cause of time-translation symmetry breaking in this system \cite{vK2016b}.  This spectral pairing is equivalent to $\Delta \mathcal{P}^{\omega}=1/2.$ The time crystal phase can thus be viewed as arising from a nonzero quantized non-adiabatic geometric effect on the Floquet lattice. We illustrate the probability densities of the spectral pairs for both phases in Fig.\ref{Fig_Majorana}c,d. The general case with interactions can be studied as well, although it requires a discussion of many-body localization which we will carry out elsewhere. 

\begin{figure}
\includegraphics[trim = 0mm 0.cm 0cm 0mm, clip,scale=0.14]{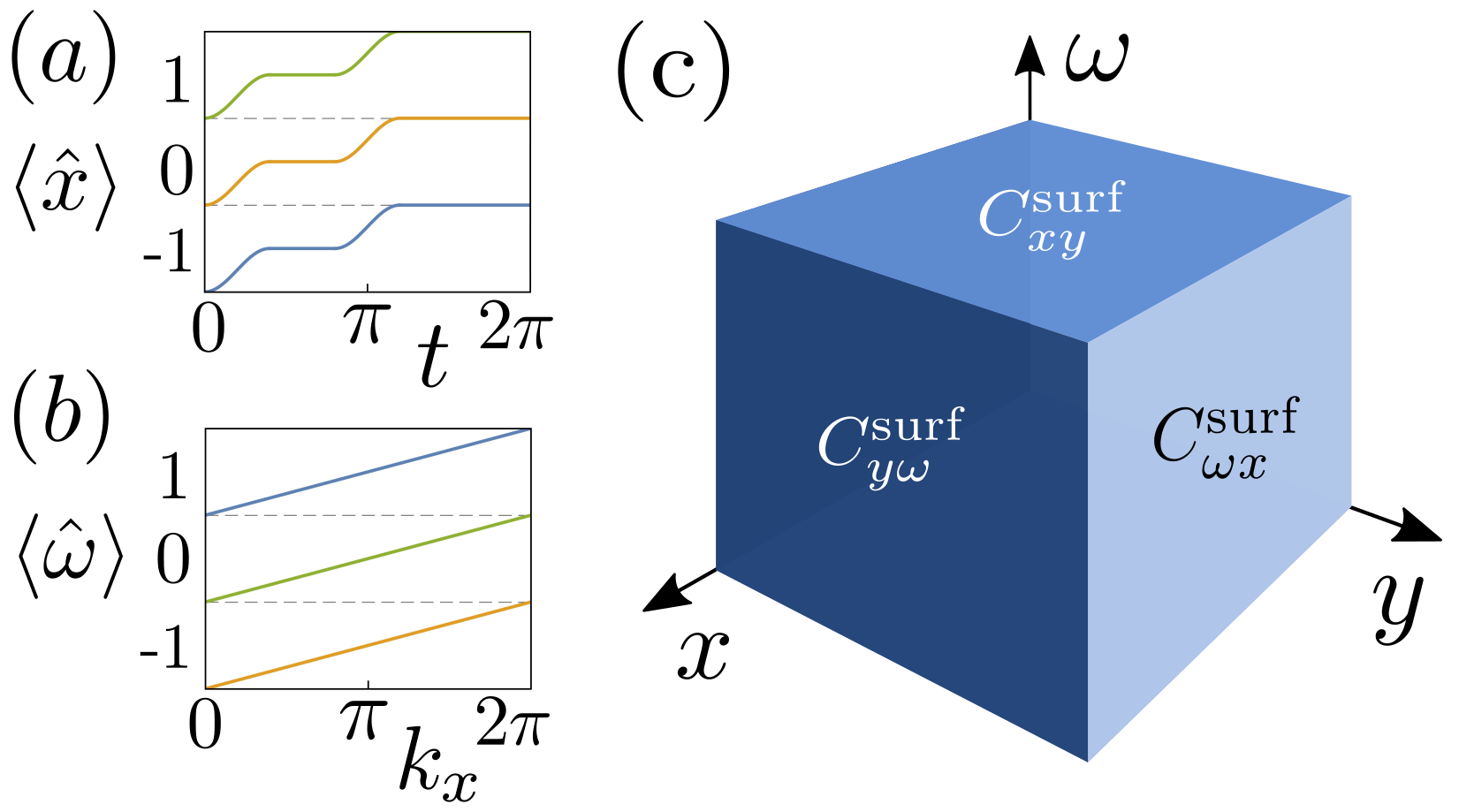}
\caption{ Pumping of charge (\textbf{a}) and photons (\textbf{b}) at the surface $y=N_y$  due to $C^{\text{surf}}_{\omega x}=1$ when $J=5/4.$ \textbf{(c)} 3D Floquet lattice with three surface Chern numbers. }\label{Fig_AFAI}
\end{figure}

\textit{Quantized frequency-domain pumping.} It is also possible to obtain quantized pumping of frequency-domain polarization in a (2+1)-dimensional system.  Consider the model $h(\mathbf{k},t)=\sum_{n=1}^4 J_n(t)\left[\cos  \left(\mathbf{b}_n\cdot \mathbf{k}\right) \sigma^1 -\sin  \left(\mathbf{b}_n\cdot \mathbf{k}\right) \sigma^2 \right],$ with $\mathbf{b}_{1}=-\mathbf{b}_{3}=\hat{x}/2,$  and $\mathbf{b}_{2}=-\mathbf{b}_{4}=\hat{y}/2,$ and where $\sigma^a$ are Pauli matrices that act on orbitals $\{A,B\}$ in which $\sigma^3$ is diagonal \cite{Titum2016}. The coefficient $J_n(t)= J $ only if $(n-1)T/5<t<nT/5,$ ($n=1\ldots 4$), and vanishes otherwise.  The system in this case resides on a 3D Floquet lattice  (Fig.\ref{Fig_AFAI}), and the phase diagram includes a topological phase that has chiral modes at the boundary and vanishing Chern number in the bulk, which we will focus on. 

Consider first the analytically solvable point $J=5/4$ with open boundary conditions along the $\hat{y}$ direction. A surface mode $\ket{u^{R}_{0 k_x}}=\ket{N_y,A}$ will appear that is localized at $y=N_y$ and has weight only on the $A$ orbital. This mode has a quasi-energy spectrum given by $e^{-2\pi i \epsilon_{R,k_x}}=e^{- i\left(k_x -\pi\right)}.$ Using $\ket{u^{R}_{tk_x}}=\mathcal{U}_{k_x}(t,0)\ket{N_y,A},$ one finds that the average energy per cycle vanishes and, as a consequence, its frequency-domain polarization is given by $e^{-2\pi i \mathcal{P}^\omega_{R,k_x}}=e^{- i\left(k_x -\pi\right)}.$ Remarkably, as $k_x$ changes by $2\pi,$  $\mathcal{P}^\omega_{R,k_x}$ increases by an integer, which signals the presence of a Chern number at this surface. Indeed, by calculating the Berry curvature $\mathcal{F}_{\omega x}(k_x, t)=\partial_{t} \mathcal{A}_{k_x}-\partial_{k_x} \mathcal{A}_{t},$ we obtain $C^{\text{surf}}_{\omega x}=\frac{1}{2\pi}\int_0^{2\pi}\int_0^{2\pi} dk_x dt \mathcal{F}_{\omega x}(k_x, t)=\int^{2\pi}_0 dk_x \partial_{k_x}\mathcal{P}^{\omega}_{k_x}=1.$ A similar calculation at the opposite surface $y=1$ leads to $e^{-2\pi i \mathcal{P}^\omega_{R,k_x}}=e^{- i\left(-k_x +\pi\right)},$ implying that at that surface $C^{\text{surf}}_{\omega x}=-1.$ 

These surface Chern numbers are a robust property of this phase of matter when the bulk is localized by disorder, which is known as the anomalous Floquet-Anderson insulator (AFAI) \cite{Titum2016}. Let us again focus on the $\omega x$ surface. Consider threading flux $\theta_x$ parallel to the $\hat{y}$ direction. Since the bulk is localized, only the delocalized surface modes are sensitive to flux insertion. The number of boundary modes is given by $ \mathcal{N}_{\text{edge}}=\sum_j\int_0^{2\pi} d \theta_x \partial_{\theta_x}  \epsilon_{j\theta_x}$ \cite{Titum2016}, where $j$ runs over states localized in a vicinity of one of boundaries of the system. Using $\epsilon_{j\theta_x}=\overline{E}_{j\theta_x}+\mathcal{P}^{\omega}_{j\theta_x},$ we obtain
\begin{eqnarray}
\mathcal{N}_{\text{edge}}=\frac{1}{2\pi}\int_0^{2\pi}\int_0^{2\pi} d\theta_x dt \mathcal{F}_{\omega x}(\theta_x, t)=C^{\text{surf}}_{\omega x},
\end{eqnarray}
where we have used that $\bra{u^{(j)}_{t\theta_x}}h(\theta_x,t)\ket{u^{(j)}_{t\theta_x}}$ is a smooth and periodic function of $\theta_x,$ so that $\int^{2\pi}_0 d\theta_x \partial_{\theta_x}\overline{E}_{j\theta_x}=0$.  Since the $y\omega$ surface has the same number of modes, then $C^{\text{surf}}_{y\omega}=C^{\text{surf}}_{\omega x}.$  

There are two physically complementary ways to interpret a nonzero Chern number \cite{Laughlin1981}, which we illustrate when $J=5/4$ in Fig.\ref{Fig_AFAI}.  From one perspective, a nonzero $C^{\text{surf}}_{\omega x}$ implies that there is quantized charge pumping in the $\hat{x}$ direction as $t$ traverses a full period (Fig.\ref{Fig_AFAI}a), which was pointed out using different arguments in [\onlinecite{Titum2016}]. The second point of view, which emerges from our study of frequency-domain polarization, is that the surface states move in the $\hat{\omega}$ direction by $ C^{\text{surf}}_{x\omega}$ unit cells as $k_x$ traverses the BZ (Fig.\ref{Fig_AFAI}b). This generalizes in the disordered case to the sum of displacements along the frequency axis of all edge states, adding up to $ C^{\text{surf}}_{x\omega},$  as a flux quantum is threaded.   Heuristically, if all edge states are occupied, we can interpret this net displacement as a number $\vert C^{\text{surf}}_{x\omega}\vert$ of photons being emitted (when $C^{\text{surf}}_{x\omega}>0$) at one edge of the system  as a flux quantum is threaded; the opposite edge correspondingly absorbs the same number of photons. We thus find that the AFAI is not only a non-adiabatic charge pump, but it is also a quantized photon pump. 

Finally, by truncating the frequency domain, surface Chern numbers $C^{\text{surf}}_{xy}$ perpendicular to the frequency direction must also arise that are equal to the number of boundary modes \cite{Rudner2013}. Thus, $C^{\text{surf}}_{xy}=C^{\text{surf}}_{\omega x}=C^{\text{surf}}_{y\omega }.$ Mechanisms for creating a boundary in the frequency domain using memory effects have been proposed recently \cite{Baum2018}.  The AFAI phase is thus characterized by Chern numbers on all of its surfaces as illustrated in Fig.\ref{Fig_AFAI}c.  

\textit{Creating new dynamical phases.} New drive-induced phases could be discovered by investigating the possible patterns of frequency-domain polarization that can be realized on the Floquet lattice. In practice, we may exploit our knowledge of ground states with nontrivial polarization in $d$ spatial dimensions to construct, by analogy, Floquet phases on the $d$-dimensional Floquet lattice.

\begin{figure}
\includegraphics[trim = 0mm 0.cm 0cm 0mm, clip,scale=0.16]{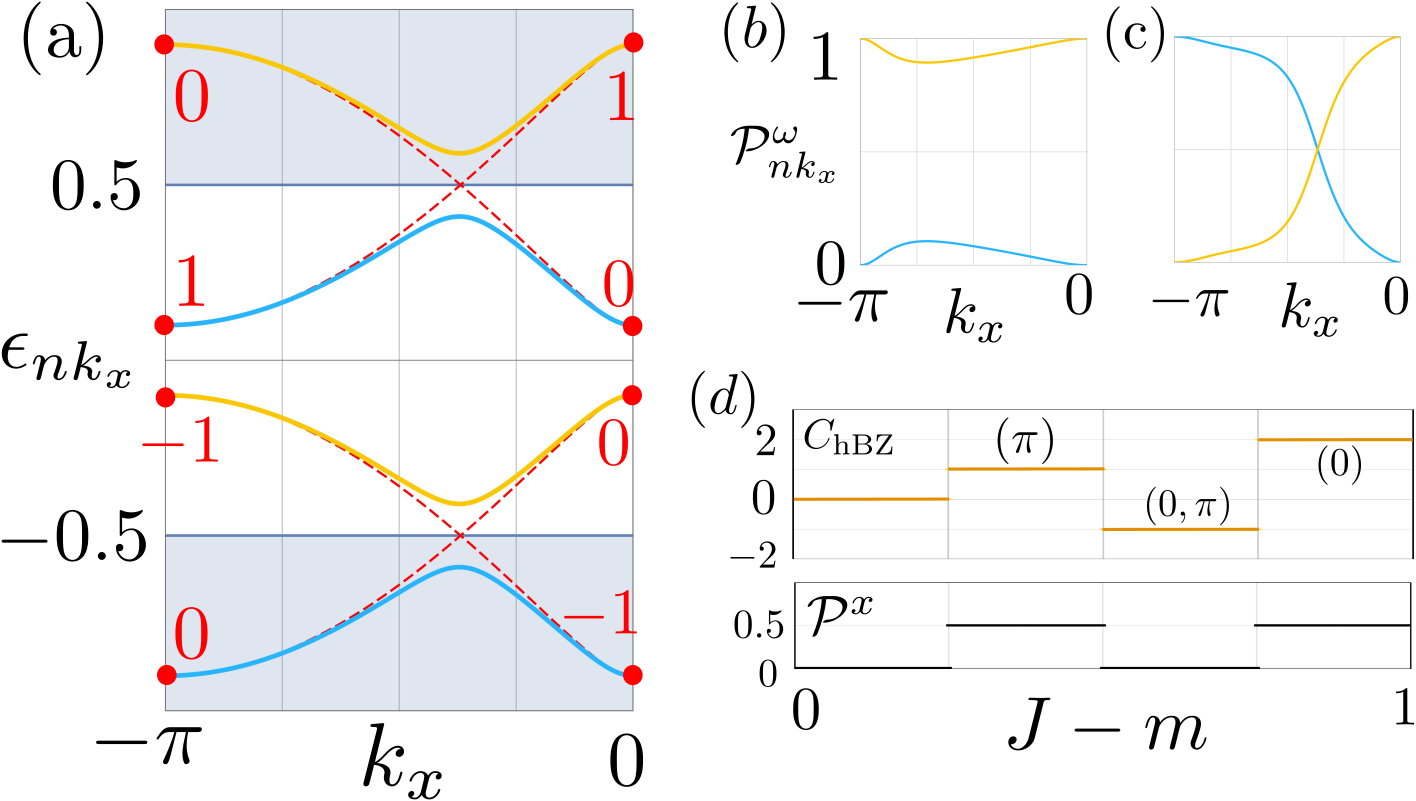}
\caption{\textbf{(a)} Quasi-energy spectrum in the repeated zone scheme. The red-dashed lines are the bands in the undriven limit. The integers indicated at $\mathcal{Q}=-\pi,0$ are the average frequencies. \textbf{(b,c)} Flow of the frequency-domain polarization for  $C^{\pm}_{\text{hBZ}}=0$ ($\lambda=0.5, m=-0.5, J=0.4$) and $ C^{\pm}_{\text{hBZ}} =\mp 1$ ($\lambda=0.5, m=-0.275, J=0.175$). \textbf{(d)} Phase diagram illustrating some Chern numbers and spatial polarization that can be obtained.}\label{Fig_NP}
\end{figure}

For example, there exist two-dimensional static topological phases that are characterized by a Chern number defined in half of the BZ (hBZ) \cite{Moore2007}. We now show that a Chern number in hBZ can also be obtained on a 2D Floquet lattice. One way to achieve this is to enforce that $\mathcal{P}^\omega_{\mathcal{Q}}$ vanish at $\mathcal{Q}=0,\pi,$ so that  $\mathcal{P}^\omega_{k_x}$ interpolates between integers as $k_x$ is varied from $0$ to $\pi.$ Consider a two-band model $h(k_x,t)=\sum^3_{a=1} d_{a}(k_x,t)\sigma^a,$ where the Pauli matrices $\sigma^a$ act on two orbitals $A,B.$ The integer constraint we seek can  be satisfied if $\mathcal{U}_{\mathcal{Q}}(t,t_0)=e^{-i \int_{t_0}^{t} d\tau d_{3}(\mathcal{Q},\tau)\sigma^3}.$ This will be the case, for example, if $h(k_x,t)$ satisfies the particle-hole symmetry $Ch(k_x,t)C^{-1}=-h^T(-k_x,t),$ with $C=\sigma^1 K$ and where $K$ represents complex conjugation. If we additionally require that $h(k_x,t)$ be a continuous function of time, a Chern number can then be defined
\begin{eqnarray}
C^{\sigma}_{\text{hBZ}}=\frac{1}{2\pi}\int^{2\pi}_0\int^{\pi}_{0} dt dk_x \mathcal{F}^{\sigma}_{k_x\omega}(k_x,t),
\end{eqnarray}
where $\sigma=\pm$ labels the two quasi-energy bands. Due to particle-hole symmetry, $C^{+}_{\text{hBZ}}=-C^{-}_{\text{hBZ}}.$ A simple numerical check reveals that the integer constraint at $\mathcal{Q}=0,\pi$ fails if one adds trivial particle-hole symmetric bands. As a result, $C^{\sigma}_{\text{hBZ}}$ is only quantized for two-band particle-hole symmetric wires, similar to static Hopf insulators \cite{Moore2008}. In addition to $C^{\sigma}_{\text{hBZ}}$, we know that particle-hole symmetric systems have a quantized spatial polarization $\mathcal{P}^{x}_n=0,1/2$ \cite{Qi2008, Budich2013}.  Thus, there exists drive-induced two-band one dimensional dynamical phases protected by particle-hole symmetry with a $\mathbb{Z}\times \mathbb{Z}_2$ classification. This is remarkable, as it goes beyond the known $\mathbb{Z}_2\times \mathbb{Z}_2$ classification of driven particle-hole symmetric topological states \cite{Roy2017}.

To construct bands with arbitrary Chern numbers, consider $h(k_x,t)=J \sin k_x \sigma^1+\left(m+J \cos k_x +\lambda \sin t \right)\sigma^3.$ We show in Fig.\ref{Fig_NP}a the spectrum in a repeated zone scheme, with two quasi-energy bands per frequency unit cell. Consider the static limit $\lambda=0$ such that the bands cross at $\epsilon=1/2$ as shown by the red dashed lines in Fig.\ref{Fig_NP}a. In this limit, the states at $\mathcal{Q}=-\pi,0$ of a given band have the same integer average frequency, as denoted next to the red dots in Fig.\ref{Fig_NP}a, producing zero Chern number. By turning on a weak drive, the quasi-energy crossings become gapped; however, the average frequency at $\mathcal{Q}$ remain fixed to integer values, as required by our construction. Due to the resulting band inversion, each band now interpolates between different integer average frequencies, leading to a nonzero Chern number. 

Through a sequence of these band inversions, arbitrary values of the Chern number can be achieved as a function of $m,J.$ In Fig.\ref{Fig_NP}c,d, we show the quantized flow of polarization for the cases $C^{\pm}_{\text{hBZ}}=0$ and $C^{\pm}_{\text{hBZ}}=\mp 1.$ The flow in the latter case is unremovable by continuous transformations of the Hamiltonian and thus indicates that the system is inherently dynamical.  In Fig.\ref{Fig_NP}d, we illustrate part of the resulting phase diagram for our model. 

Finally, the two-band constraint implies that these dynamical phases are not robust to the breaking of translational symmetry. One way to see this is that by adding a potential with a periodicity that is twice that of the original system, the unit cell is doubled. Since this doubles the number of bands in the BZ, the resulting system will not have a well-defined $C_{\text{hBZ}}.$ A consequence of this is that the bulk integer invariant is not equal to the number of boundary modes, since the boundary itself breaks translational symmetry.  Notwithstanding this, the parity of the bulk Chern number does provide some information about the presence of boundary modes. By producing band inversions as illustrated in Fig. \ref{Fig_NP}, one can infer that a band inversion at $\epsilon=1/2$ ($\epsilon=0$) always changes $C^{\pm}_{\text{hBZ}}$ by an odd (even) amount. Thus, a $\pi$-mode will arise whenever $(-1)^{i\pi C^{\sigma}_{\text{hBZ}}}=-1.$ Furthermore, each time there is a phase transition at either quasi-energy gap,  $\mathcal{P}^x$ changes by $1/2$ \cite{Budich2013}. Thus, whenever $(-1)^{i\pi(C^{\sigma}_{\text{hBZ}}+2 \mathcal{P}^x)}=-1,$ a $0$-mode will be present in the spectrum.

\textit{Conclusions.} In this work, we have studied the signatures of driven phases of matter on the Floquet lattice. Floquet eigenstates are characterized by a non-adiabatic Berry phase which we interpret as frequency-domain polarization. This polarization exhibits quantized behavior in phases induced by a drive and, thus, serves as a criterion to identify when a phase is inherently dynamical. This opens the door to understanding and discovering new driven phases with stable patterns of frequency-domain polarization.

\begin{acknowledgments}
\textit{Acknowledgements.} IMS is grateful for a discussion with C. W. von Keyserlingk and comments by Mark Rudner on the manuscript. We also gratefully acknowledge discussions with Wang Zhong. IMS and AA were supported by the Yale Prize Postdoctoral Fellowship. Work at Argonne National Laboratory was supported by the Department of Energy, Office of Science,  Materials Science and Engineering Division.
\end{acknowledgments}

\end{document}